# Discovery of a bi-critical point between antiferromagnetic and superconducting phases in pressurized single crystal $Ca_{0.73}La_{0.27}FeAs_2$

Yazhou Zhou[1]*, Shan Jiang[2]*, Qi Wu[1]*, Vladimir A. Sidorov[3], Jing Guo[1], Wei Yi[1], Shan Zhang[1], Zhe Wang[1], Honghong Wang[1], Shu Cai[1], Ke Yang[4], Sheng Jiang[4], Aiguo Li[4], Ni Ni[2], Guangming Zhang[5,6]†, Liling Sun[1,6]† & Zhongxian Zhao[1,6]

[1]*Institute of Physics and Beijing National Laboratory for Condensed Matter Physics, Chinese Academy of Sciences, Beijing 100190, China*
[2]*Department of Physics and Astronomy, UCLA, Los Angeles, CA90095, USA*
[3]*Institute for High Pressure Physics, Russian Academy of Sciences, 142190 Troitsk, Moscow, Russia*
[4]*Shanghai Synchrotron Radiation Facilities, Shanghai Institute of Applied Physics, Chinese Academy of Sciences, Shanghai 201204, China*
[5]*State Key Laboratory for Low dimensional Quantum Physics and Department of Physics, Tsinghua University, Beijing 100084, China*
[6]*Collaborative Innovation Center of Quantum Matter, Beijing, 100190, China*

One of the most strikingly universal features of the high-temperature superconductors is that the superconducting phase emerges in the close proximity of the antiferromagnetic phase, and the interplay between these two phases poses a long-standing challenge. It is commonly believed that，as the antiferromagnetic transition temperature is continuously suppressed to zero, there appears a quantum critical point, around which the existence of antiferromagnetic fluctuation is responsible for the development of the superconductivity. In contrast to this scenario, we report the discovery of a bi-critical point identified at 2.88 GPa and 26.02 K in the pressurized high-quality single crystal $Ca_{0.73}La_{0.27}FeAs_2$ by complementary *in-situ* high pressure measurements. At the critical pressure, we find that the antiferromagnetism suddenly disappears and superconductivity simultaneously emerges at almost the same temperature, and that the external magnetic field suppresses the superconducting transition temperature but hardly affects the antiferromagnetic transition temperature.

Even after thirty years from the discovery of the copper oxide superconductors, how to understand the interplay between antiferromagnetism and unconventional superconductivity has been one of the most sophisticated issues in condensed matter physics [1-9]. Iron-based superconductors found in 2008 are a new family of high temperature superconductors [10,11], which provides new opportunity to clarify this issue. Among them, the iron pnictide superconductor $Ca_{1-x}La_xFeAs_2$ [10-17] has a monoclinic structure with the -(FeAs)-(Ca/La)-As-(Ca/La)-(FeAs)- stacking along $c$ axis [15,17]. In particular, the presence of the metallic As-As zig-zag chains in the spacer layers make it structurally and electronically distinct from the architype 122 Fe based superconductor $CaFe_2As_2$ [18,19]. The superconductivity in $Ca_{1-x}La_xFeAs_2$ exhibits in the doping range from $x=0.15$ to $x=0.25$ (Ref. [15,16,20,21]). Nuclear magnetic resonance measurements found that the superconducting phase coexists with the antiferromagnetic (AFM) phase in the above doping range and the AFM transition temperature is enhanced with increasing La concentrations [22]. However, beyond the doping level ~ 0.25, the sample becomes a poor metal with a stripe like AFM long-ranged order [21], which can be regarded as the 'parent compound' of this family of superconductors. It is well-known that pressure is a 'clean' way to realize a continuous tuning of the crystal structure and the corresponding electronic structure without introducing additional chemical complexity, being an ideal method to study the interplay between the AFM and superconducting phases [23,24].

In this study, by applying a Toroid (also known as the Paris-Edinburgh-type) high-pressure cell with the glycerin/water (3:2) liquid as the pressure transmitting medium, we performed complementary *in-situ* hydrostatic pressure measurements of resistance, alternating current (*ac*) susceptibility and heat capacity on the high-quality $Ca_{0.73}La_{0.27}FeAs_2$ single crystals. The arrangement of the samples and the components on the lower part of the pressure anvil is shown in Fig.1A. Under ambient pressure a resistivity drop at the temperature ~54 K is observed (Fig.1B), signifying an AFM transition ($T_M$) that is in fairly agreement with the reported results from neutron scattering measurements [21]. Above $T_M$, the first derivative of the resistance with respective to the temperature shows a weak kink which is associated with the monoclinic to a triclinic structural phase transition (inset of Fig.1B), indicating that the stripe-like AFM is characterized by three-component order parameters [21]. The temperature of the structural phase transition is suppressed by applying pressure and gradually loses the resolution from the AFM transition. Under given pressures, the temperature dependence of electrical resistivity is displayed in Fig.1C, in which the evolution of $T_M$ with external pressure is confirmed by the other experimental run (Fig. S1, Supplementary Information). Our two independent measurements for two samples consistently demonstrate that the $T_M$ shifts to lower temperature upon increasing pressure. At the pressure 2.85 GPa, the AFM phase transition occurs at 26.08 K, but suddenly disappears when pressure is higher than it. Furthermore, our high-pressure heat capacity measurements verify the above results and the detailed analysis is given in Fig. S2, Supplementary Information.

With further increasing pressure just greater than 2.85 GPa, a pressure-induced superconducting phase with a transition temperature ($T_C$) of 26.09 K is found at 2.95 GPa (sample A). The values of $T_C$ are determined by the onset temperatures of the resistivity drops. The zero resistant superconducting state is observed at 3.45 GPa and above (inset of Fig.2A). To characterize the superconducting transition further, we applied magnetic field at 2.95 GPa and 4.95 GPa, respectively. It is found that the onset-temperatures of the superconducting transition shift to lower temperatures (Fig.2B), indicating that the resistivity drop is associated with a superconducting transition. To further confirm the pressure-induced superconductivity, we performed *in-situ* high pressure resistivity and alternating-current (*ac*) susceptibility measurements for another sample (sample B). The results are shown in Fig. 2C and 2D, where the $T_C$ (*ac*) is determined by the intersection of the lines through the steep slope and the zero slope. By comparing the amplitude of diamagnetic throw of the sample with that of the pressure gauge Pb (employed as a reference, placed next to the sample in the same coil, as shown in Fig.1A) which is with the similar shape and mass to the sample, we can estimate the relative change of the superconducting volume, *i.e.* from ~ 60 % at pressure of 2.99 GPa to ~97% at 3.53 GPa (Inset of Fig.2D), implying that the pressure-induced superconductivity is abruptly turned on at the critical pressure. Because the Toroid high pressure cell can only maintain the hydrostatic pressure conditions for the measurement as high as ~5 GPa [25], we performed our measurements below this pressure.

Our high-pressure x-ray diffraction measurements demonstrate that no pressure-induced crystal structural phase transition occurs throughout the pressure range investigated (Fig. S3, Supplementary Information). Thus it can be regarded that the pressure-induced suppression of AFM and emergence of superconductivity are caused by the electron-electron interactions. Our data of heat capacity versus pressure obtained at different temperatures are presented in Fig.3A-3D, respectively. These data indicate that, upon cooling, clear heat capacity jumps at the critical pressure Pc can be observed below the critical temperature $T_{BC}$=26.02 K. These jumps, defined as $\Delta C$, directly reflect a first-order phase transition from the AFM phase to the superconducting phase. Remarkably, $\Delta C$ is found to vary with temperature (Fig.3E). As the temperature decreases down to the $T_{BC}$, $\Delta C$ suddenly appears. Below $T_{BC}$, $\Delta C$ continuously reduces until it is undetectable at the temperatures lower than 8 K. Such a feature of the changes in the $\Delta C$ indicates a first-order phase transition.

Next, we summarize the pressure dependence of the characteristic temperatures $T_M$ and $T_C$ in the phase diagram of Figure 4A, in which there are two distinct low temperature regions representing the AFM phase and superconducting phase. In the AFM phase region, the $T_M$ is remarkably suppressed with increasing pressure and terminated at 26.08 K and 2.85 GPa. Then the superconductivity with $T_C$ = 26.09 K appears at 2.95 GPa, where the $T_C$ is determined from the electrical resistance measurements. The values of the $T_C$ are unexpectedly close to the terminating point of $T_M$ (26.08 K) at 2.85 GPa, which implies that the AFM transition temperature and the superconducting transition temperature will meet together at a critical pressure. Extrapolations of the $T_M$-$P$ curve and $T_C$-$P$ curve yield an intersected point at 2.88 GPa and 26.02 K, which is denoted by the red star in the phase diagram. Such a special point is conventionally defined as a bi-critical point in phase transition theory. In the superconducting phase region, the $T_C$ shows a weak response to pressure. Our deduced phase diagram is different from that obtained by the chemical doping in $Ca_{1-x}La_xFeAs_2$, whose superconducting phase coexists with the AFM phase [22,26].

To further understand the nature of the bi-critical point, we applied different magnetic fields perpendicular to (100) plane of the sample since an external magnetic field usually has little effect on the $T_M$ but has an effective influence on the $T_C$. We find, upon increasing magnetic field, the $T_C$ is suppressed to lower temperature in the pressure range of the superconducting phase (Fig.4B to 4D), suggestive of a common feature of field-induced suppression of superconductivity. However, $T_M$ has no visible change to the applied magnetic field under the magnetic field up to 8T (Fig.4B-4D and Fig. S4, Supplementary Information). Our results reveal that the magnetic field can make the bi-critical point no longer existed, resulting in a similar pressure–temperature phase diagram to that of the pressurized $Ca_{10}(Pt_3As_8)(Fe_2As_2)_5$ [27]. For our $Ca_{0.73}La_{0.27}FeAs_2$ compound at the critical pressure, the value of $T_C$ is suppressed by 18.5% at 8T, compared with the value at zero-field. The extrapolation of the $T_C$ value suggests that the requested magnetic field to suppress $T_C$ to 0 K is about 44T (Fig.S5, Supplementary Information).

To the best of our knowledge, this is the first time that a bi-critical point between an AFM phase and a superconducting phase is discovered experimentally in high temperature superconductors. Twenty years ago, the SO(5) theory attempted to unify antiferromagnetism and superconductivity in the temperature-doping phase diagram of copper oxide superconductors and predicted a bi-critical point with emergent SO(5) symmetry (Ref. [28,29]). If this higher symmetric bi-critical point exists, it can be expected that the application of magnetic field will not be able to separate the intersected point of the $T_M$ and $T_C$. The separation of $T_M$ and $T_C$ at the critical pressure observed throughout our measurements implies that the SO(5) theory is not applicable to interpret the bi-critical point found in the pressurized $Ca_{0.73}La_{0.27}FeAs_2$. Therefore, our discovery of the bi-critical point provides a unique experimental foundation for understanding the interplay between the AFM and superconductivity, and is expected to pave a path to finalize the debate on the mechanism of high Tc superconductivity.

**Acknowledgements**

The work was supported by the NSF of China (Grants No. 91321207, No. 11427805, No. U1532267, No. 11404384), the Strategic Priority Research Program (B) of the Chinese Academy of Sciences (Grant No. XDB07020300), the National Key Research and Development Program of China (Grant No.2016YFA0300300), the Russian Foundation for Basic Research (Grant No. 15-02-02040) and the U. S. NSF DMREF (DMR-1435672).

† Correspondence and requests for materials should be addressed to L.S. (llsun@iphy.ac.cn) and G.M.Z. (gmzhang@tsinghua.edu.cn)

*These authors are contributed equally.

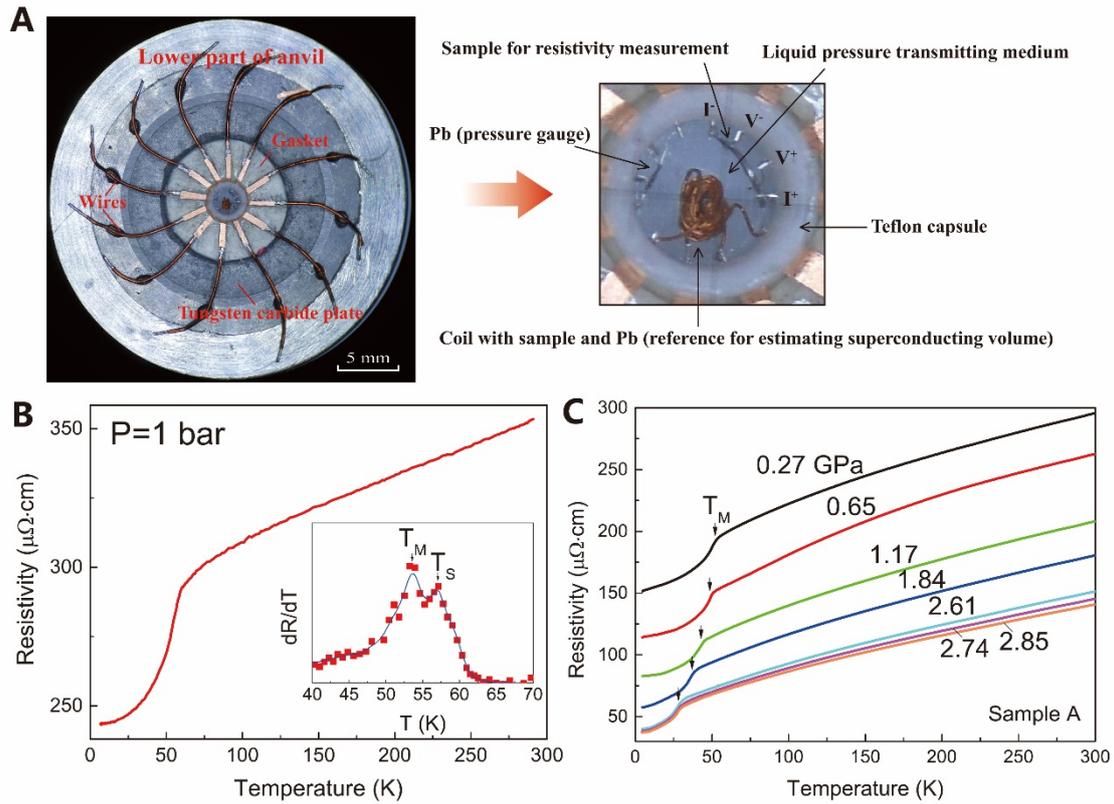

Figure 1 Arrangement of samples, measurement components in a Toriod high pressure cell, and determinations of the antiferromagnetic transition temperature ($T_M$) in pressurized single crystal $Ca_{0.73}La_{0.27}FeAs_2$ through the electrical resistance measurements. **(A)** Top review of the arrangement of the samples and components for electrical resistance and *ac* susceptibility measurements on the lower part of the pressure anvil. (B) Temperature dependence of the resistivity in the sample A at 1 bar. Inset displays the derivative of resistivity with respective of temperature, and $T_S$ represents the temperature of structure phase transition. (C) Resistivity as a function of temperature in the sample A for pressures ranging from 0.27 GPa to 2.85 GPa. The arrows in figure (C) indicate the AFM transition temperature $T_M$, which shows a decrease with pressure.

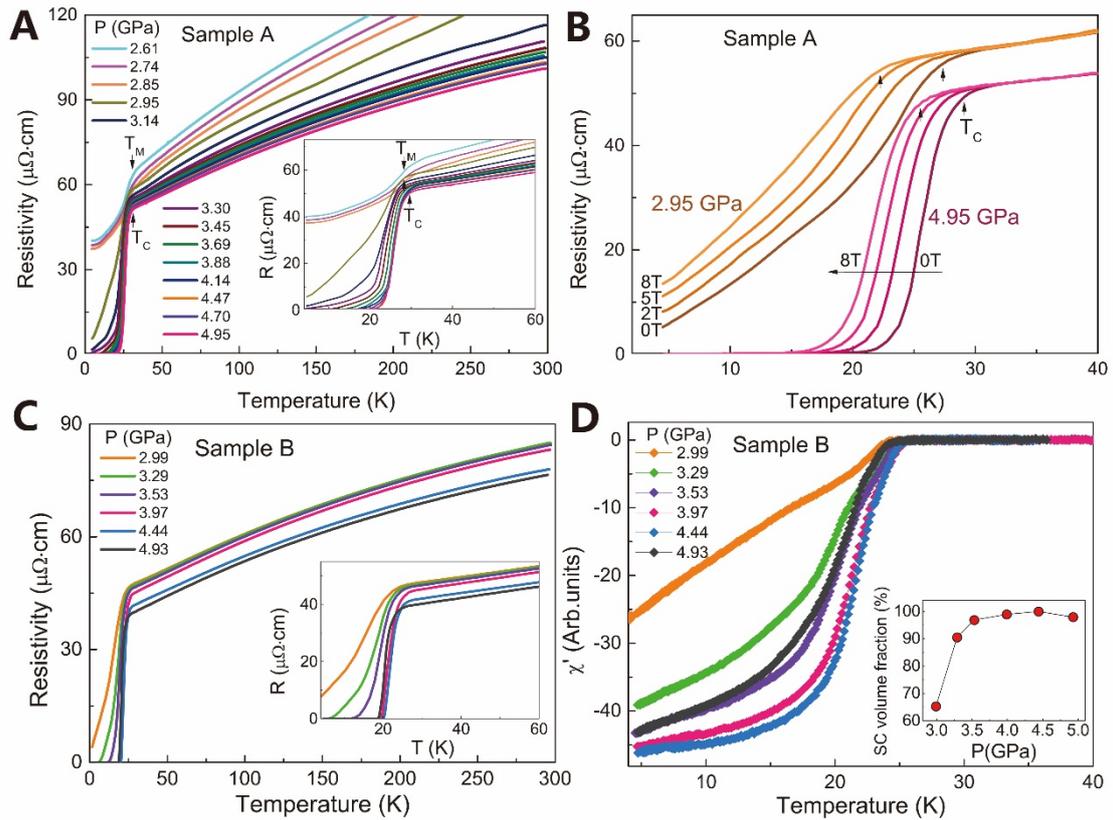

Figure 2 Characterizations of pressure-induced superconductivity in $Ca_{0.73}La_{0.27}FeAs_2$. (A) Resistivity as a function of temperature in the sample A for pressures ranging from 2.61 GPa to 4.95 GPa, displaying an evolution from AFM to superconducting transition. Inset illustrates the enlarged view of resistivity around the critical temperatures. (B) Plots of the resistivity-temperature curves of the sample A under different magnetic fields for selected pressures of 2.95 GPa and 4.95 GPa. (C) Temperature dependence of the electrical resistivity in the sample B with further increasing pressures, showing a superconducting transition starting at 2.99 GPa. The low-temperature resistivity is zoomed in for a clearer view (inset). (D) Real part of the *ac* susceptibility as a function of temperature in the sample B at different pressures, clearly demonstrating the diamagnetic signals. The inset shows the pressure versus

estimated superconducting (SC) volume.

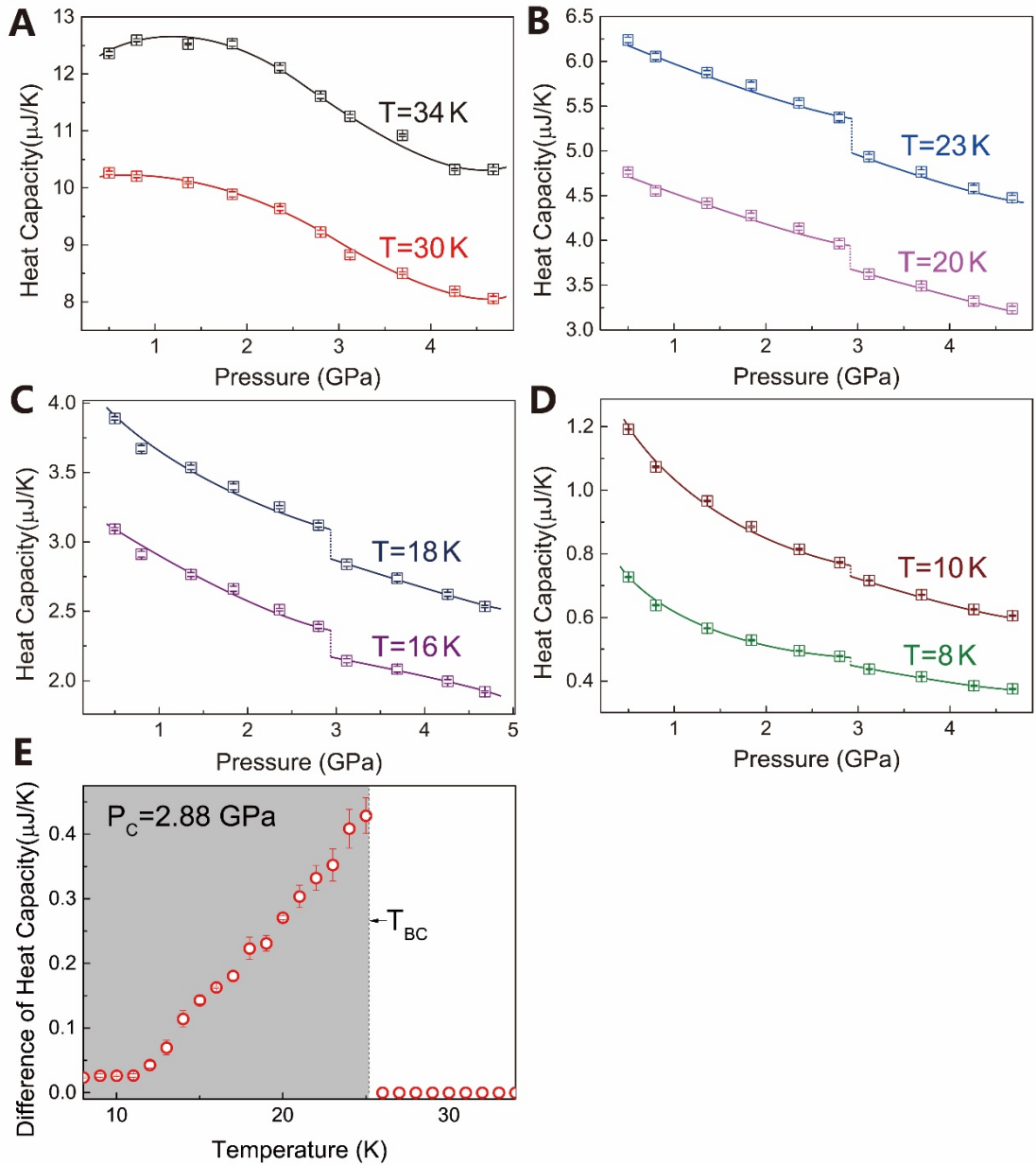

Figure 3 High pressure heat capacity data. (A) Pressure dependence of the heat capacity for given temperatures showing the evolution from AFM phase to PM phase. (B)-(D) Heat capacity as a function of pressure at given temeparures, showing jumps at the crossover from the AFM phase to superconducting phase below the bi-critical

point temperature ($T_{BC}$). (E) The jump value ($\Delta C$) of the heat capacity in figure (B-D) as a function of temperature, illustrating that the $\Delta C$ reaches a maximum at the temperature of the bi-critical point.

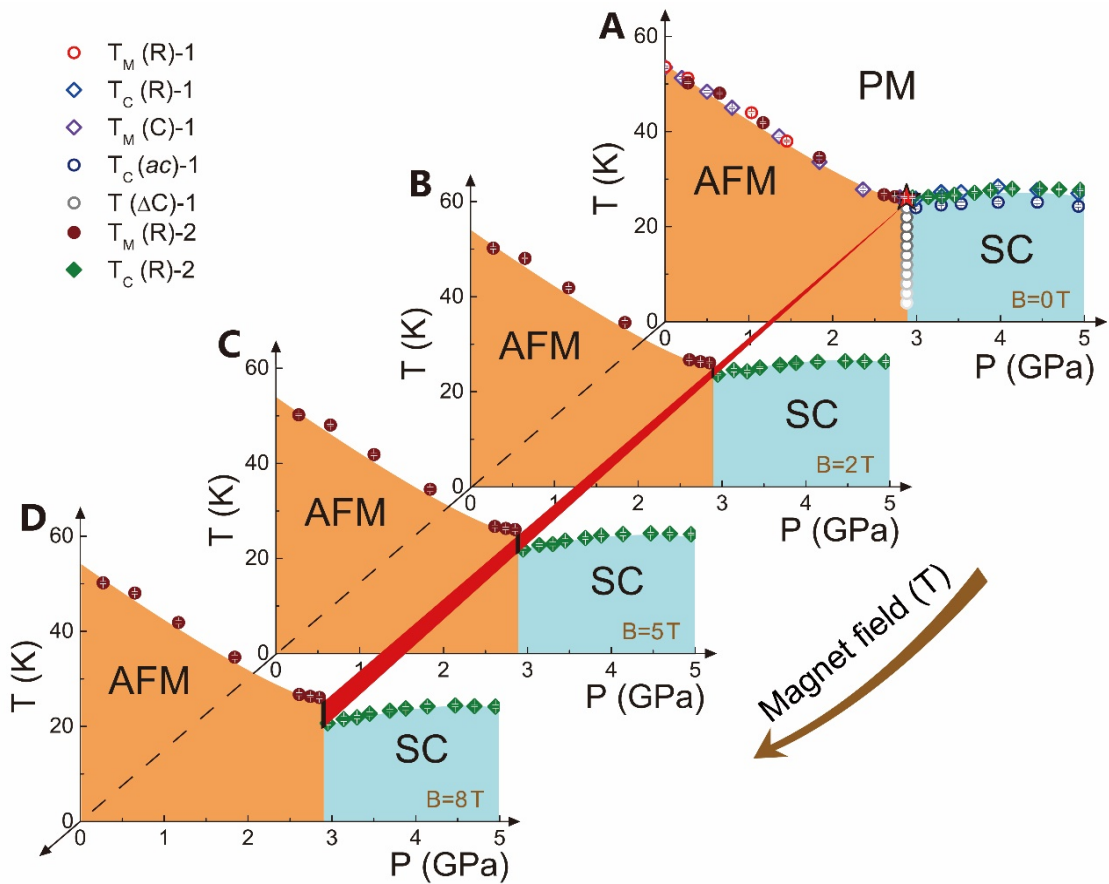

Figure 4 Temperature-pressure phase diagrams obtained at different magnetic fields for $Ca_{0.73}La_{0.27}FeAs_2$ single crystals. The acronym PM, AFM and SC stand for paramagnetic, antiferromagnetic and superconducting phases, respectively. The wine circles and custom diamonds represent the temperature of the AMF phase transition detected from two-run electrical resistance measurements ($T_M(R-1)$, $T_M(R-2)$) and heat capacity $T_M(C)$ measurements under hydrostatic pressure condition. The green diamonds and blue circles stand for the superconducting transition temperature

determined from the resistance $T_C$(R-1) and *ac* susceptibility $T_C$(*ac*) measurements, respectively. The position of red star denotes the location of the bi-critical point, which is determined by an intersection of extrapolated lines of the pressure dependent $T_M$ and $T_C$. The gray circles are the data extracted from the heat capacity results. The radial red line shows the suppressed tendency of the $T_C$ by magnetic field.